\documentstyle[12pt,epsfig]{article}
\newcommand{\bq}{\begin{equation}}
\newcommand{\eq}{\end{equation}}
\newcommand{\bqn}{\begin{eqnarray}}
\newcommand{\eqn}{\end{eqnarray}}
\newcommand{\nb}{\nonumber}
\newcommand{\lb}{\label}
\def\R{{\bf{R}}}
\def\noi{\noindent}

\begin{document}

\title{Gravitational Collapse of Self-Similar and Shear-free Fluid
with Heat Flow }
\author{R. Chan$^1$, M. F. A. da Silva$^2$, and Jaime F. Villas da
Rocha $^2$ \\
\small \it $^1$ Coordenadoria de Astronomia e Astrof\'{\i}sica,
Observat\'orio
Nacional \\
\small \it Rua General Jos\'e Cristino 77, S\~ao Crist\'ov\~ao,
$20921-400$ \\
\small \it Rio de Janeiro, RJ, Brazil \\
\small\it $^2$ Departamento de F\' {\i}sica Te\' orica,
Universidade do Estado do Rio de Janeiro, \\
\small\it Rua S\~ ao Francisco Xavier $524$, Maracan\~ a,
$20550-013 $ \\
\small \it Rio de Janeiro, RJ, Brazil }

\maketitle

\begin{abstract}

A class of solutions to Einstein field equations is studied, which
represents gravitational collapse of thick spherical shells made
of self-similar and shear-free fluid with heat flow. It is shown
that such shells satisfy all the energy conditions, and the
corresponding collapse always forms naked singularities.

\end{abstract}

\newpage
%

\section{Introduction}

One of the most outstanding problems in gravitation theory is the final
state
of a collapsing massive star, after it has exhausted its nuclear fuel.
Despite
numerous efforts over the last three decades, our understanding is still
limited to several conjectures, such as, the cosmic censorship conjecture
\cite{Penrose}, and the hoop conjecture \cite{Thorne}. To the former,   many
counter-examples have been found \cite{Joshi}, although  it is still not
clear
whether  those particular examples are stable and generic. To the latter,
no counter-examples have been found so far in four-dimensional Einstein's
Theory of gravity, although it has been shown
recently that this is no longer the case in five dimensions \cite{NM01}.

Lately,  by studying gravitational collapse of spherically symmetric dust
fluid, Joshi, Dadhich and Maartens (JDM) \cite{JDM01}
present some results that lead them to the conclusion that the
formation of naked singularities is due to shear of the fluid.
So, sufficiently
strong shearing effects delay the formation of apparent horizons, thereby
exposing the strong gravitational regions to the outside world and leading
to
naked singularities. This is an important and unexpected result, because
from
the Raychaudhuri equation,
one can see that the shear contributes positively to the focusing
effect \cite{Ray55}. On the other hand, it is well-known that pressures also
play a very  significant role in gravitational collapse. Thus, it is
natural to ask  what kind of roles that shear might play in models with
non-vanishing pressures. As pointed out in Ref. \cite{JDM01}, when pressures
are present the  problem becomes very complicated, and it is still not clear
which role that  shear might play in these models.
However, there is at least, as far as we know, two counter-examples
to the JDM conjecture \cite{deutsche, ind02},
both of them are particular spherically symmetric models with heat flow
(although these authors do not emphasize this relevance of their results).
Note also that in both cases
their solutions are  conformally
flat.

In this paper, we shall study a class of solutions, which
represents gravitational collapse of a conformally flat,
shear-free and self-similar fluid with heat flow. For this class
of solutions, we shall show that the collapse always forms naked
singularities. Similar results were also found in
\cite{deutsche,ind02}. Gravitational collapse of fluid with
self-similarity has been intensively studied recently \cite{CC98}.
In addition, gravitational collapse of fluid with heat flow has
been also studied lately \cite{WS01}.

The rest of the paper is organized as follows: in Section 2 we
shall present the solutions found by Som and Santos (SS) some
years ago \cite{SS81}, which represents a shear-free fluid with
heat flow. We shall show that some  of these solutions have
self-similarity of the zeroth, first and second  kinds, according
to the classifications given by Carter and Henriksen \cite{CH89}.
In the study of a fluid with heat flow, one of the difficulties is
to verify the energy conditions. Although the problem is simple
but often involved very tedious calculations, and sometimes one
has to study them numerically \cite{deutsche}. Thus, we  devote
Section 3 to study the plausible physical conditions
\cite{HE73,Kolassis88}. In the Sections 4 and 5 we present the
junction conditions for an exterior Vaidya's spacetime \cite{Vai}
and for an interior Minkowski spacetime. Such constructed
solutions can can be considered as representing gravitational
collapse of spherical thick shells. It is shown that the collapse
always forms naked singularities.    The paper is closed by
Section 6, in which our main conclusions are presented.


\section{Self-Similar and Shear-free Fluid with Heat Flow}

%

\begin{figure}[]
\begin{center}
\leavevmode
\psfig{file=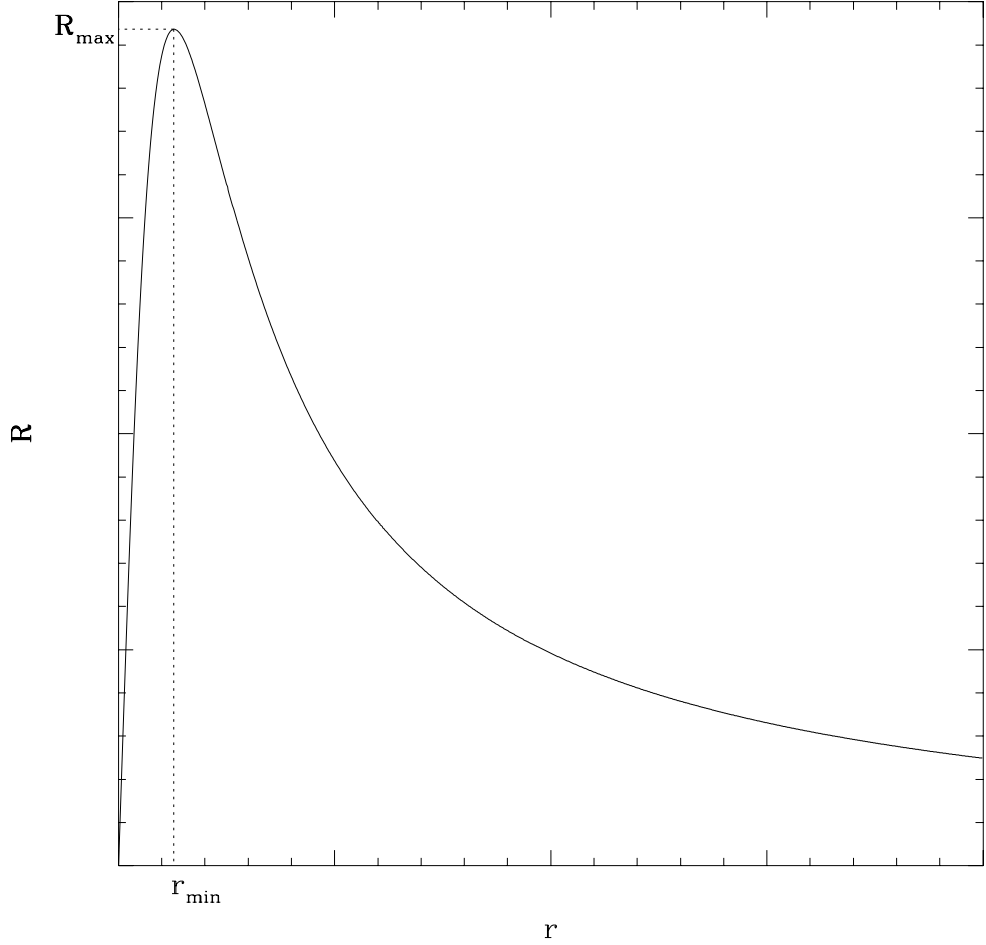,width=1.0\textwidth,angle=0}
\caption{The geometrical radius $\R$ as a function of the coordinate $r$.
For $r$ $\leq$ $r_{min}$ , we have $r$ $=$
$(1-\sqrt{1- 4 f_1 f_2 {\R} ^2})(2 f_1
{\R} ^2)^{-1}$ and $r$ $\geq$ $r_{min}$, we have $r$ $=$
$(1+\sqrt{1- 4 f_1 f_2 {\R} ^2})(2 f_1
{\R} ^2)^{-1}$ where $r_{min}$ $=$  $\sqrt{f_2/f_1}$. The maximal
geometrical radius, $\R_{max}$ = $\R(r_{min})$, is given by
$(-t)/(2\sqrt{c})$.}
\label{fig1} \end{center}
\end{figure}

In this section, we shall study the solutions found by Som and
Santos \cite{SS81}, which represent shear-free anisotropic fluid
with heat flow. The SS solutions are given by
  \bq
  \lb{2.1}
  ds^2  = A^2(t,r) \left(dt^2 - dr^2 - r^2 d\Omega^2\right),
  \eq
where $ d\Omega^2 \equiv d\theta^2 + \sin^2(\theta) d\phi^2,\;
x^\mu$ = $\{t,r,\theta,\phi\}$, and
 \bq
 \lb{2.1a}
 A(t, r) =
\frac{1}{f_1(t)r^2+f_2(t)},
 \eq
with $f_{1}(t)$ and $f_{2}(t)$ being arbitrary functions. This
solution has an unusual behavior for the geometrical radius,
defined as
 \bq
 \lb{2.1b} \R = \frac{r}{f_1(t)r^2+f_2(t)}.
 \eq
\noi As a matter of fact,   when  $r$ $\rightarrow$ $0$ we have
$\R$ $\rightarrow$ $0$. In addition, if $f_{1}(t)f_{2}(t) > 0$, we
have $\R$ $\rightarrow$ $0$, too, as $r$ $\rightarrow$ $\infty$.
Then, at any given moment $t = t_{1}$, we can see that $\R$ is
always bounded, as can be seen from the figure (\ref{fig1}).

The corresponding energy-momentum tensor (EMT) takes the
form\footnote{In this paper we shall choose units such that the
Einstein coupling constant $\kappa$ is unity.},
 \bq
\lb{2.1c}
T_{\mu\nu} \; = \rho u_{\mu}u_{\nu} - {\tilde p}h_{\mu\nu} +
q_{\mu}u_{\nu} + q_{\mu}u_{\nu},
 \eq
where $u_{\nu}$ is the four-velocity of the fluid, $\rho$ the
energy density measured by comoving observers with the fluid,
$h_{\mu\nu} $ denotes the inducted metric on the three-surfaces
orthogonal to $u_{\nu}$, and $\tilde p$ is the effective pressure.
In terms of the  isotropic pressure $p$ and  the second
coefficient of the viscosity, $\zeta$,  $\tilde p$ is given by
$\tilde p =  p - \zeta {u^\mu}_{;_\mu}$. In equation (\ref{2.1c})
the heat flow vector $q^{\mu}$ satisfies the condition $q^\mu
u_\mu$ = 0 and in the comoving coordinates is given by \bq
\lb{2.1e} q^{\mu} =  - \frac{4r}{k A^2}
{f_{1}'(t)}\delta^{\mu}_{r}, \eq where a prime denotes the
ordinary differentiation with respect to $t$. For the details, we
refer readers to Ref. \cite{SS81}.

In this paper, we shall consider a particular case of the SS  solutions
(\ref{2.1a}), where the fluid is not viscous
and the corresponding spacetime has self-similarity.

Generalizing the concepts of Newtonian Mechanics \cite{BZ72} to General
Relativity, Carter and Henriksen (CH)
\cite{CH89} gave the notion of {\em kinematic
self-similarity} with its
properties,
\bq
\lb{2.6}
{\cal{L}}_{\xi}h_{\mu\nu} = 2 h_{\mu\nu},\;\;\;\;
{\cal{L}}_{\xi}u^{\mu} = -\alpha u^{\mu},
\eq
where ${\cal{L}}_\xi$ is a Lie operator,
$h_{\mu \nu}$ is the project operator, defined by
$h_{\mu \nu}$ = $g_{\mu \nu}$ - $u_{\mu} u_{\nu}$,
and $\alpha$ is an arbitrary constant. When $\alpha = 1$, it can be shown
that
the kinematic self-similarity reduces to the {\em homothetic} one (or
self-similarity of {\em the first kind}), which
was first studied by Cahill and Taub in GR for a perfect fluid \cite{CT71}.
When $\alpha \not= 1$, CH argued that this would be a natural relativistic
counterpart of self-similarity of {\em the second kind} ($\alpha \not= 1$)
and
of {\em the zeroth kind} ($\alpha = 0$) in Newtonian Mechanics.

Applying the conditions (\ref{2.6})  to the metric (\ref{2.1}), it can be
shown that the metric coefficient $A(t,r)$  has to take the form,
\bq
\lb{2.3c}
A(t, r) = A(z),
\eq
where in each case the conform Killing vector $\xi^{\mu}$ and the
self-similar
variable $z$ are given by
\bq
\lb{1.12}
\xi^{\mu}\frac{\partial}{\partial x^{\mu}} = \frac{\partial}{\partial t} +
r \frac{\partial}{\partial r},\;\;\;
z = {r}e^{-t},\; (\alpha = 0),
\eq
for the zeroth kind, and
\bq \lb{1.11}
\xi^{\mu}\frac{\partial}{\partial x^{\mu}} = \alpha
t\frac{\partial}{\partial
t} + r \frac{\partial}{\partial r},\;\;\;
z = \frac{r}{\left(- t\right)^{1/\alpha}},\; (\alpha \not= 0),
\eq
for the second kind. In the latter case when $\alpha = 1$ it degenerates to
the first kind. For the details, we  refer to \cite{CH89,Coley97,fred02}.

\subsection{Solutions with Self-similarity of the Zeroth Kind}

>From Eqs.(\ref{1.12}) and (\ref{2.3c}),  we can see that for the SS
solutions
(\ref{2.1a}), the  ones   with  self-similarity of the zeroth kind
are given by
\bq
\lb{2.zc1}
f_1(t) = e^{-2t}, \; \; \; \; \;   f_2 (t) = c,
\eq
where $c$ is an arbitrary constant. The spacetime singularities are
manifested
from the associated Kretschmann scalar,
\bqn
\lb{2.zc2}
{\cal{K}} &=& R^{\mu \nu \lambda \sigma}R_{\mu \nu \lambda \sigma}
  =  -16 e^{-8t} \left\{
\left(12r^4-28r^{2}+15\right)c^{2} e^{-4t}
\right. \nb\\
& &  \left.  + r^2\left[ c e^{2t}\left( 4 r^{2} -6\right)
+r^2\left(12r^4 -16r^2 +3\right) \right] \right\},
 \eqn
from which we can see that it is singular at $ t$ = -$\infty$ and
$r$ = $\infty$ $(\R = 0)$. It can be shown that in this case the
solutions cannot be interpreted as representing gravitational
collapse. Thus, in the following we shall not consider them any
more.

\subsection{Solutions with Self-similarity of the First and Second Kinds}

The SS solutions with self-similarities of the first and
second kinds   are given by
\bq
\lb{2.sc1}
f_1(t) = (-t)^{-2/\alpha} \; \; \; \; \;
  f_2 (t) = c,
\eq
where $c$, as in the last case, is a constant.
Then, it can be shown that the corresponding Kretschmann scalar reads
\bqn
\lb{2.zc3}
{\cal{K}}  &=& -\frac{16}{\alpha^{4} (-t)^{4(2+\alpha)/\alpha}}
\left\{c^2 ({-t})^{4/\alpha}
\left[15(\alpha t)^4 \right.\right. \nb\\
& & \left. - 2 \alpha^3(3\alpha +14) t^2 r^2
+3(\alpha^2 + 2)^2 r^4 \right]\nb\\
& &
+ c r^2({-t})^{2/\alpha} \left[ 6 \alpha^2 r^4 - 6(\alpha t)^4
+4 t^2 r^2 +12 \alpha r^4 \right] \nb\\
& & \left. + r^4\left[3(\alpha t)^4 + 2\alpha^2(3\alpha -8) r^2
t^2 +3(\alpha^2+4) r^4 \right] \right\}, \eqn which shows that the
spacetime now is singular  at $t = 0$.

The apparent horizon of the fluid can be defined as the outer hypersurface
that satisfies the equation \cite{Poisson}

\begin{figure}[]
\begin{center}
\leavevmode
\psfig{file=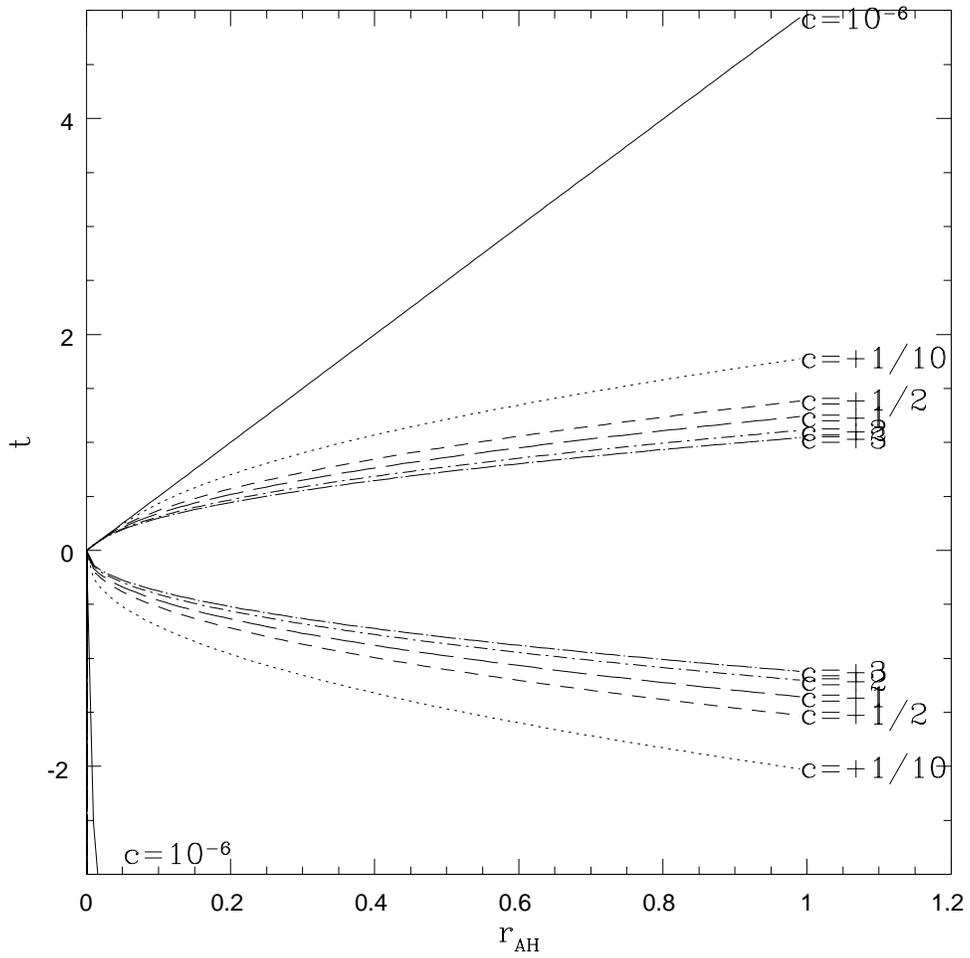,width=1.0\textwidth,angle=0}
\caption{The apparent horizon radius for $\alpha=2/5$.}
\label{fig2}
\end{center}
\end{figure}

\begin{figure}[]
\begin{center}
\leavevmode
\psfig{file=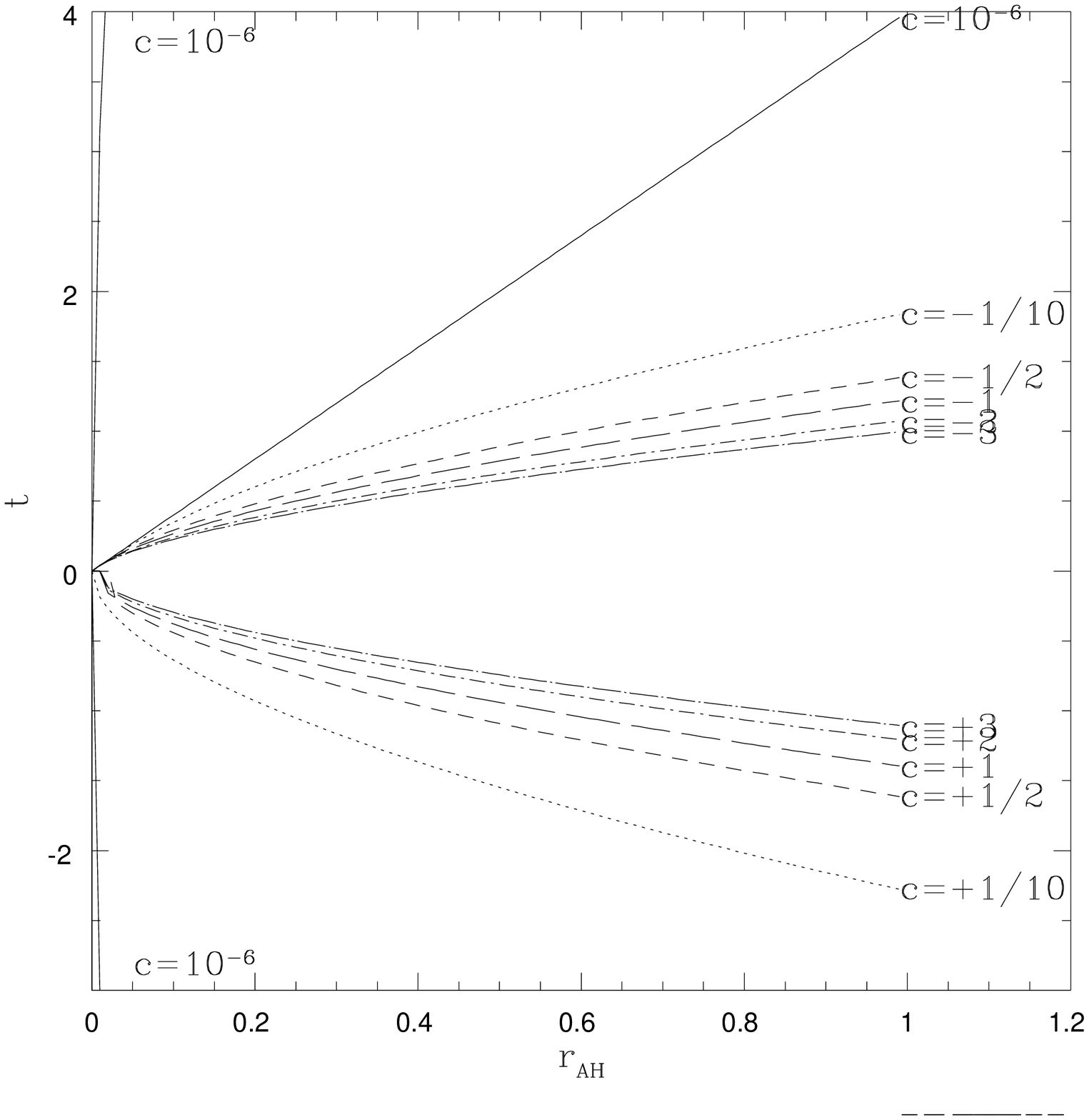,width=1.0\textwidth,angle=0}
\caption{The apparent horizon radius for $\alpha=1/2$.}
\label{fig3}
\end{center}
\end{figure}

\begin{figure}[]
\begin{center}
\leavevmode
\psfig{file=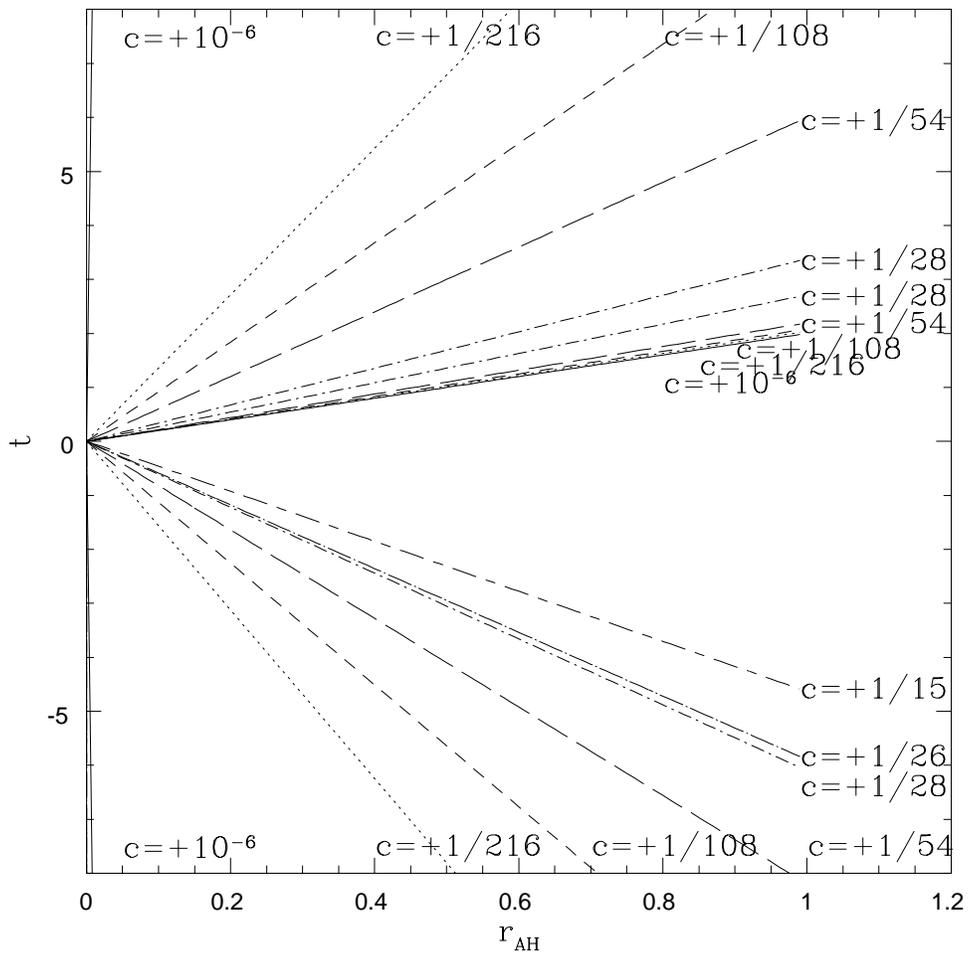,width=1.0\textwidth,angle=0}
\caption{The apparent horizon radius for $\alpha=1$.}
\label{fig4}
\end{center}
\end{figure}

\bq
\lb{AH}
\R_{,\alpha}\R_{,\beta}g^{\alpha\beta}=0.
\eq
It will be shown below that the apparent horizon of the fluid is always
outside of the three energy condition frontiers. Besides, 
from  equation (\ref{2.1b}),
we can see that the geometrical radius vanishes when $t$ = 0.
Figures (\ref{fig2}), (\ref{fig3}) and (\ref{fig4}) show the solutions of 
equation (\ref{AH}) for various values of $c$ and $\alpha$.
Their comparison, and the fact that the case with 
$\alpha$ = 1 (first kind of self-similarity) can
be considered as a sub-case of the second kind
lead to the conclusion 
that the cases with first and second kind of self-similarities 
present the same physics. So,  in the following, 
we will study only the solutions with
first kind.

It should be noted that to
consider these solutions as physical, they have to satisfy some physical and
geometrical conditions \cite{HE73, Kolassis88}. We shall devote the next
section to study these conditions. Before proceeding, we would also like to
note that the spacetimes of the above solutions are singular at $z =
\infty$, too, as one can see from Eqs.(\ref{2.zc2}) and
(\ref{2.zc3}). However, as we shall show below, the energy
conditions will restrict these solutions valid only in the region
$ z_{1} < z < z_{2}$, where $z_{1,2}$ are finite constants. Thus,
the singularity at $z = \infty$ becomes irrelevant in the models
to be constructed below.

\eject
\section{Plausible Physical Conditions}
Although the solution (\ref{2.1a}), at first sight, is mathematically
very simple, in virtue of the difficulties imposed by (\ref{2.1b}),
we need to  enterprise a very careful analysis, in order to
assure their correct interpretation and, with this, select  those
subsets of solutions that can represent astrophysical collapse.

\subsection{Energy Conditions}
Since the corresponding EMT has heat flow and, as a
result, it is not diagonal,  then the analysis of the energy conditions
becomes considerably complicated.
In order to study
these energy conditions, one has first to cost it in its canonical form
\cite{HE73}. We find that  solutions with self-similarity of the first kind
have similar properties as those with self-similarity of the second and
zeroth
kinds. Thus,  in the following we shall consider only the solutions with
$\alpha  = 1$. For these solutions it can be shown that the   EMT given by
equation (\ref{2.1c}) can be written as,
\begin{equation}
\lb{3.15}
T_{\mu\nu}=\rho
t_{\mu}t_{\nu}+ q (t_{\mu}r_{\nu}+r_{\mu}t_{\nu})
+ p \left(\theta_{\mu}\theta_{\nu}+ \phi_{\mu}\phi_{\nu}\right),
\end{equation}
where $t_{\mu}= A\delta^t_{\mu},\; r_{\mu}= A\delta^r_{\mu}$, $\;
\theta_{\mu}=r A\delta^\theta_{\mu}$, $\;
\phi_{\mu}=rA\sin{\theta}\delta^\phi_{\mu}$,
and
\bqn
\lb{3.16}
\rho &=& {12\over t^2}(c+z^4),\;\;\;
q = -{8z\over t^2}(c + z^2)^2,\nb\\
p  &=&    {4\over t^2}[-2c + (1+3c)z^2],
\eqn
\noindent where $z$ = $r/(-t)$, as defined in equation (\ref{1.11}),
is the self-similar variable.
To write the  EMT in its canonical form, we need to solve the eigenvalue
problem,
\bq
\lb{4.1}
\tau^{\mu}_{\nu} \xi^{\nu} = \lambda \xi^{\mu},
\eq
which  will possess nontrivial solutions only
when the determinant ${\rm det}|\tau^{\mu}_{\nu} - \lambda
\delta^{\mu}_{\nu}|
= 0$ that can be written as \cite{TW00},
\bq
\lb{4.2}
\left(\lambda + p\right)^{2}\left[\left(\lambda-\rho\right)
\left(\lambda +p\right)+ q^2\right] = 0.
\eq
Clearly, the above equation has four roots,
$\lambda_{1,2} = p$ and $\lambda _{\pm}$,
where
\bq
\lb{4.4}
\lambda _{\pm} = {(\rho-p)\over 2} \pm \Delta,\;\;\;
\Delta^2 \equiv \frac{(\rho + p)^{2} - 4 q^{2}}{4} .
\eq
The eigenvalues $\lambda_{1,2}$
correspond to, respectively,
the eigenvector $\xi_{2}^{\mu} = \theta^{\mu}$ and $\xi_{3}^{\mu} =
\phi^{\mu}$, which represent the two principal  tangent
directions. On the other hand, substituting equation (\ref{4.4}) into
equation ({\ref{4.1}), we find that the corresponding eigenvectors are given
by
\bq
\lb{4.5} \xi^{\mu}_{\pm} =
(\lambda_{\pm} + p)t^{\mu} + q r^{\mu}.
\eq

The conditions  $\Delta^2 > 0,\; \Delta^2 = 0$ and $\; \Delta^2 < 0$
divide the whole spacetime into several regions. In the following let us
first consider regions where $\Delta^2 > 0$.

\subsubsection{Regions with $\Delta^2 > 0$}

>From equation (\ref{4.4})  we find that
\bq
\lb{4.9}
\Delta^2 = {4\over t^4}(9z^2-1)(z^2-1)(z^2+c)^2.
\eq
Thus,  the regions where $\Delta^2 > 0$ depend on the values of the constant
$c$. In particular, we find that
\[
\Delta^2 > 0 \;\;\; =>
\]
\bq
\lb{4.9a} =>
\cases{
z < 1/3, \; {\rm or} \; z > 1,
                     & $  c > 0$,\cr
z < \sqrt{-c}  \; {\rm or} \; z >1
\; {\rm or} \; \sqrt{-c} < z < 1/3 \; {\rm or} \;
1/3 < z <1 ,
                      & $ -1/9 < c < 0$, \cr
z < 1/9 \; {\rm or} \; z> 1 \; {\rm or } \;
1/3 < z < \sqrt{- c} \; {\rm or}\;  \sqrt{-c} < z < 1 ,
                      & $  -1 < c < - 1/9,$ \cr
z < 1/9  \; {\rm or} \; z > \sqrt{-c} \; {\rm or} \;
1/3 < z < 1 \; {\rm or} \; 1 < z < \sqrt{-c} \;
                      & $ c < -1 $. }
\eq
As can be seen from equation (\ref{4.4}), now the two
roots $\lambda _{\pm}$ and the two eigenvectors $\xi_{\pm}^{\mu}$ are all
real
and satisfy the relations,
\bqn
\lb{4.10}
(\lambda_{+} + p)(\lambda_{-} + p) &=& q^{2},\nb \\
\frac{\xi^{\mu}_{\pm} \xi^{\nu}_{\pm} g_{\mu\nu} }
{\Delta(\lambda_{\pm} + p)} &=& \pm 1,\nb\\
\xi^{\mu}_{+} \xi^{\nu}_{-} g_{\mu\nu} &=& 0.
\eqn
>From these expressions we can see that when $\lambda_{+} + p > 0$,
the eigenvector $\xi_{+}^{\mu}$ is timelike, and $\xi_{-}^{\mu}$
is spacelike, while when $\lambda_{+} + p < 0$,
the two vectors exchange their roles.

{\bf Case A.1) $\; \lambda_{+} + p > 0$}: This condition can be written as
\bq
\lb{4.10a}
\rho+p+2\Delta > 0.
\eq
Setting
\bqn
\lb{4.11}
E_{(0)}^{\mu} &\equiv& \frac{\xi^{\mu}_{+}}{\left[D^{1/2}(\lambda_{+} +
p)\right]^{1/2}},\nb\\
E_{(1)}^{\mu} &\equiv& \frac{\xi^{\mu}_{-}}{\left[D^{1/2}(\lambda_{-} +
p)\right]^{1/2}},\nb\\
 E_{(2)}^{\mu} &\equiv& \theta^{\mu},
E_{(3)}^{\mu} \equiv \phi^{\mu},
\eqn
we find that $E_{(a)} ^{\mu},\; (a = 0,1,2,3)$ form an orthogonal basis,
i.e., $E^{\lambda}_{(a)}E_{(b)\lambda} = \eta_{ab}$, with $\eta_{ab} = {\rm
diag.}\{1,\; -1, \; -1,\; -1\}$.
Then, in terms of these unit vectors, the  EMT given by
equation (\ref{3.15}) takes the form

\bq
\lb{4.14}
\left(T_{(a)(b)}\right) =
\left(\matrix{
\rho_{(0)} & 0 & 0 & 0 \cr
0 & p_{(1)}  &0 & 0 \cr
0 & 0  & p_{(2)} & 0 \cr
0 & 0  &0 & p_{(3)} \cr }\right),
\eq
where $T_{(a)(b)} \equiv T_{\mu\nu}E_{(a)}^{\mu}E_{(b)}^{\nu}$, and
\bqn
\lb{4.13}
\rho_{(0)} &=& {1\over 2}(\rho-p+2\Delta),\nb\\
p_{(1)} &=& {1\over 2}(p-\rho+2\Delta),\nb\\
p_{(2)} &=& p_{(3)} = p.
\eqn
equation (\ref{4.14}) corresponds to the Type I fluid defined in
\cite{HE73}.

Once the EMT is cast in its canonical form, we can apply the energy
conditions to it. A simple but tedious analysis of these conditions
reveals that all of them are satisfied only for the case where
$ 0 < c < 1/15$ and
$z_{1} < z < z_{2}$   as showed in Fig. \ref{fig6}. The hypersurfaces
$z = z_{1,2}$ are determined as follows:

\begin{figure}[]
\begin{center}
\leavevmode
\psfig{file=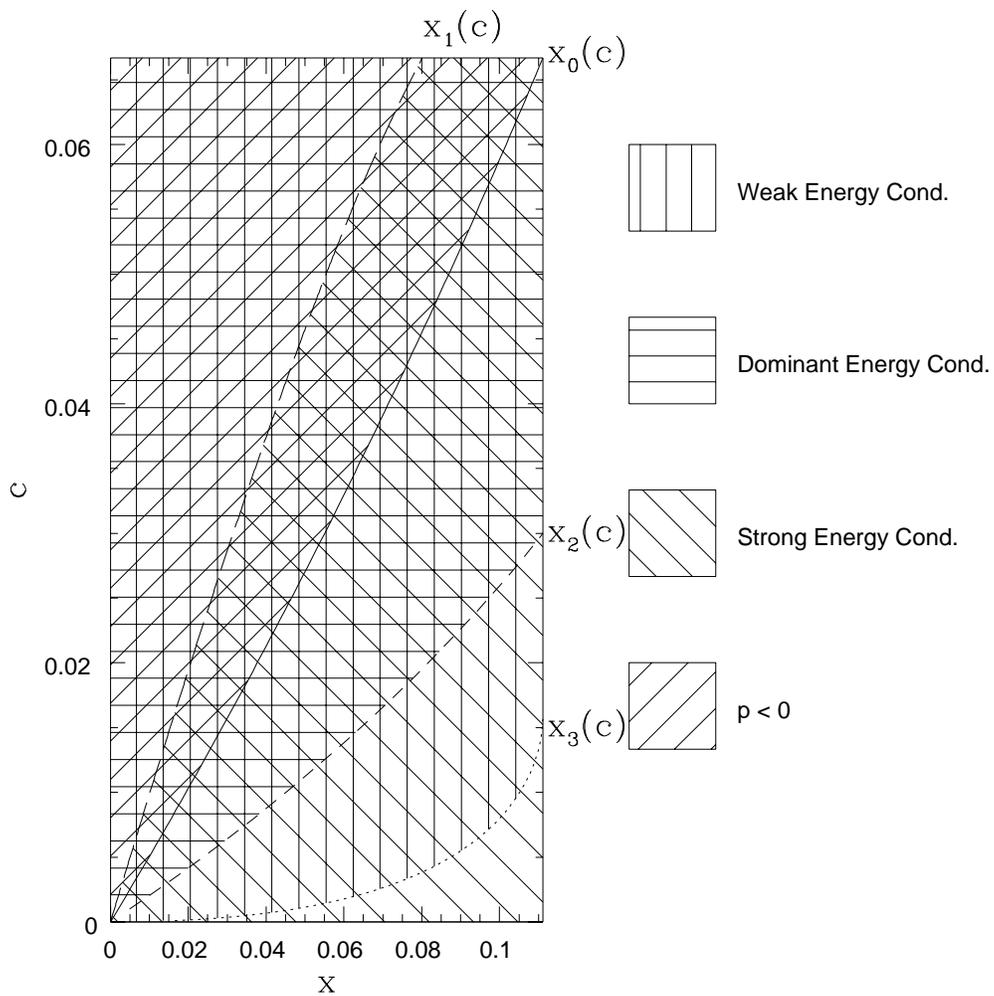,width=1.0\textwidth,angle=0}
\caption{The different intervals
where the energy conditions
are satisfied. For the range $\rm{x}$ between ${\rm{x}}_1$
and ${\rm{x}}_2$ where $\rm{x}$ = $z^2$ all the three energy conditions
are satisfied}
\label{fig6}
\end{center}
\end{figure}

The boundary $ z = z_{1}$ comes from the condition   $\rho_{(0)}+p_{(1)}+2p
\geq 0$, and is given by the first positive root of the   equation
\bq
9x^4+2(9c-5)x^3-(3+44c+27c^2)x^2+(18+38c)cx-15c^2=0,
\eq
while the boundary $z = z_2$ is obtained from the condition $\rho_{(0)}-p
\geq
0$ and given by the second positive root of the   equation
\bq
(1+9c)x^3-[(1+3c)^2+10c]x^2+c[7(1+3c)+2c]x-10c^2=0,
\eq
where $x \equiv z^{2}$. In between these two boundaries, all the rest of
energy
conditions are identically satisfied.

{\bf Case A.2) $\; \lambda_{+} + p < 0$}: This condition can be written as
\bq
\lb{4.20}
\rho+p+2\Delta < 0.
\eq
Since now $\xi^{\mu}_{-}$ is time-like, the orthogonal basis can be chosen
as
\bq
\lb{4.17}
E_{(a)}^{\mu} \equiv\left\{\frac{\xi^{\mu}_{-}}{D^{1/4}\left|\lambda_{-} +
p\right|^{1/2}},\frac{\xi^{\mu}_{+}}{D^{1/4}\left|\lambda_{+} +
p\right|^{1/2}},\theta^{\mu}, \phi^{\mu}\right\}.
\eq
Then, it can be shown that the corresponding EMT also takes the form of
equation (\ref{4.14}) but now with
\bqn
\lb{4.18}
\rho_{(0)} &=& {1\over 2}(\rho-p-2\Delta),\nb\\
p_{(1)} &=& {1\over 2}(p-\rho-2\Delta),\nb\\
p_{(2)} &=&  p_{(3)} = p.
\eqn
>From equation (\ref{4.20}) one can show that the weak and dominant energy
conditions are violated for any given values of $c$. Thus, in
the following discussions we shall discard this case.

Summarizing the results of this subsection we can see that the
fluid satisfies all the three energy conditions only in the region where
$z_{1} < z < z_{2}$ or equivalently, $ r_{1}(t) < r < r_{2}(t)$, where
$r_{1}(t) = - z_{1} t$ and $r_{2}(t) = - z_{2} t$.

\subsubsection{The Hypersurfaces where $\Delta^2=0$}
>From equation (\ref{4.9}) we can see that now $\Delta^2=0$ represents the
hypersurfaces\footnote{Since in this paper we are mainly concerned with
gravitational collapse of the  fluid, we consider only the region
where $t \le 0$ or $z \ge 0$.} $z = 1/3,\; 1,\; \sqrt{- c}$,   on which we
have
\bq
q=-{1\over 2}(\rho+p),
\eq
and the two roots $\lambda_{\pm}$ given by equation (\ref{4.4})
degenerate into one. As shown in \cite{TW00}, this multiple root
corresponds to two null independent eigenvectors,
\bq
\lb{4.23}
\xi^{\mu}_{\pm} = \frac{ u^{\mu} \pm X^{ \mu } }{ \sqrt{2} }.
\eq
>From these two null vectors we
can construct two unit vectors, one is timelike and the other is spacelike,
but these are exactly $u^{\mu}$ and $r^{\mu}$. Thus,
in the basis
\bq
\lb{4.32}
E_{(a)}^{\mu} = \left\{u^{\mu},\; r^{\mu},\; \theta^{\mu},\;
\phi^{\mu}\right\},
\eq
the  EMT takes the form
\bq
\lb{4.33}
\left(T_{(a)(b)}\right) =
\left(\matrix{
\rho & q & 0 & 0 \cr
q& p & 0 & 0\cr
0& 0 & p &0 \cr
0& 0& 0 &p \cr}
\right).
\eq

To consider the energy conditions on these hypersurfaces, it is found
convenient to  distinguish the three cases $q > 0,\; q = 0$ and $q < 0$.

{\bf Case B.1)} $\; q > 0$:  In this case,   the corresponding  EMT
(\ref{4.33}) can be written in
the form
\bq
\lb{4.34}
\left(T_{(a)(b)}\right) = q
\left(\matrix{
1 + \kappa  &1 & 0 & 0 \cr
1 &1-\kappa & 0 & 0\cr
0 &0 & p_{(2)} &0 \cr
0 &0 &0 & p_{(3)} \cr}
\right), \;\;\; (q> 0),
\eq
where
\bq
\lb{4.24}
\kappa \equiv \frac{\rho - p}{\rho + p},\;\;\;\;
p_{(2)} = p_{(3)}\equiv \frac{2p}{\rho + p}.
\eq
equation (\ref{4.34}) is exactly in the form of the type
$II$ fluid classified in
\cite{HE73}. Applying the three energy conditions to this case we find that
all of them are satisfied for $z=1$, $c < -1$ and
$z=1/3$, $c < 1/63$.

{\bf Case B.2)} $\; q=0$: In this case it can be shown that  the EMT becomes
diagonal and  the energy conditions are these given for the EMT of
equation (\ref{4.14}). Then, it can be shown that only on the hypersurface
$z = 1$
with $c = -1$ all the three energy conditions are satisfied.

{\bf Case B.3)} $\; q < 0$: In this case, it can be shown that
the corresponding  EMT cannot be written in the form
of equation (\ref{4.34}). In order to study the energy conditions, let us
consider
an observer with its four-velocity given by
\bq
\lb{4.27a}
w^{\mu} = \alpha t^{\mu} + \beta r^{\mu} + \gamma \theta^{\mu}
+ \delta \phi^{\mu},
\eq
where $\alpha,\; \beta,\;\gamma$ and $\delta$ are arbitrary constants,
subject
to the condition,
\bq
\lb{4.28a}
w^{\mu}w_{\mu} = \alpha^{2} - \beta^{2} -\gamma^{2} - \delta^{2}
\ge 0.
\eq
The weak energy condition requires that \cite{HE73}
\bq
\lb{4.29a}
T_{\mu\nu} w^{\mu}w^{\nu} = \alpha^{2}\rho + \gamma^{2} p
+ \delta^{2}p - 2 \alpha\beta q \ge 0.
\eq
It can be shown that equation (\ref{4.29a})
is satisfied for any observer given by
Eqs.(\ref{4.27a}) and (\ref{4.28a}) only when the conditions
$\rho \ge 0$, $\rho + p \ge 0$, $\rho + p - 2q\ge 0 $ and $\rho
+ p + 2q\ge 0$ are true. On the other hand, the strong energy
condition holds when \cite{HE73}
\bqn
\lb{4.30a}
\left(T_{\mu\nu} - \frac{1}{2}g_{\mu\nu}T\right)w^{\mu}w^{\nu}
&=& \frac{1}{2}\left[(\alpha^{2} + \beta^{2} + \gamma^{2} +
\delta^{2})\rho \right.\nb\\
& & + (\alpha^{2} + \gamma^{2} - \beta^{2} -
\delta^{2})p\nb\\
& &  \left. + (\alpha^{2} + \delta^{2} - \beta^{2}
- \gamma^{2})p - 4 \alpha \beta q\right] \ge 0,
\eqn
which is equivalent to $\rho + p \ge 0$, $ \rho + p - 2q \ge
0$, $ \rho + p + 2q\ge 0 $ and $\rho + p + p \ge 0$. Meanwhile,
the dominant energy condition requires that   $\rho
\ge |p|$, $ \rho \ge |p|$ and $ \rho \ge |q|$. To summarize,
for any given $T_{\mu\nu}$ of the form (\ref{4.33}), the energy conditions
are
the following:

\begin{description}

\item (a){\em The Weak Energy Condition}:
\bq
\lb{4.31a}
i) \; \rho \ge 0,\;\;\;
ii)\; \rho + p \ge 0,\;\;\;
iii)\; \rho + p + 2q \ge 0,\;\;\;
iv)\; \rho + p - 2q \ge 0.
\eq

\item (b) {\em The Dominant Energy Condition}:
\bq
\lb{4.31b}
i) \;\; \rho \ge \left|p\right|,\;\;\;
ii) \;\; \rho \ge \left|p\right|,\;\;\;
iii) \;\; \rho \ge \left|q\right|.
\eq

\item (c) {\em The Strong Energy Condition}:
\bqn
\lb{4.31c}
& & i) \;\; \rho \ge 0,\;\;\;
ii) \;\; \rho + p \ge 0,\;\;\;
iii) \;\; \rho + p - 2q \ge 0,\nb\\
& &
iv) \;\; \rho + p + 2q \ge 0,\;\;\;
v) \;\; \rho + 2p \ge 0.
\eqn

\end{description}

Applying the above energy conditions to the  fluid, we find that it
satisfies all of them only on the hypersurfaces
 $z = 1$ with $c \geq -1$
and $z=1/3$ with $1/18  \leq c \leq 7/9 $.

In summary, in the present case we have only isolated
hypersurfaces satisfying all the energy conditions.

\subsubsection{Regions with $\Delta^2 < 0$}

>From Eqs.(\ref{4.4}) and (\ref{4.5}),   we can see that now the
eigenvalues $\lambda_{\pm}$ are complex, and so do the two
eigenvectors $\xi^{\mu}_{\pm}$. This means that in the present
case the EMT cannot be diagonalized (by real similarity
transformations). Then, in terms of the four unit vectors
$\{u^{\mu}$, $\; r^{\mu}$, $\; \theta^{\mu}$, $\;  \phi^{\mu}\}$,
it will take exactly the same form as that given by equation
(\ref{4.33}) but now with $q$ being arbitrary. The analysis of the
three energy conditions will follow precisely the same steps as we
did in the last subcase for $q < 0$, so  the three energy
conditions are also those given
 by Eqs.(\ref{4.31a})-(\ref{4.31c}).

It is not difficult to show that in this case none of these three
energy conditions is satisfied. In fact, from the condition
$\Delta^{2} < 0$ we find that
 \bq
\label{4.40}
(\rho+p)^{2}
-4q^{2}  < 0,
 \eq
while   the   weak energy and strong conditions all require
$(\rho+p)^{2} -4q^{2} \ge 0$. Since the dominant energy condition
is stronger than the weak one, we can see that in the present
case, all the three energy conditions are violated.

Thus, from the above analysis   we can see that, in order to interpret  the
self-similar  fluid given by Eqs.(\ref{2.zc1}) and (\ref{2.sc1})
as representing gravitational collapse, the solutions need to be restricted
to
certain regions. For the homothetic case, we have shown that it is
given by $z_{1} < z < z_{2}$ and $- \infty < t \le 0$, where $ z_{1} < z_{2}
<
1/3$. Outside of this regions, the spacetime will be described by other
solutions. For example, if we consider the hypersurface $z = z_{1}$, or $ r
=
r_{1}(t)$ then we may match it to a
 Schwarzschild, Vaidya,
or any region of other spacetimes for $r < r_{1}(t)$ or \R  $>$ $\R_1(t)$.
On the other hand, inside of the
fluid $r > r_{2}(t)$ or \R  $<$ $\R_2(t)$,
the spacetime can be matched to
Minkowski spacetime. Since the hypersurfaces $z = z_{1,2}$ or $r
= r_{1,2}(t)$ are time-like, in principle these matchings are always
possible.
Then, the resultant spacetime will represent the collapse of a shell of
a fluid with finite thickness. The spacetime is free of
singularities at initial, but due to the collapse of the fluid, a spacetime
singularity will be finally formed at the moment $t = 0$ at the origin
$\R = 0$. This singularity
is naked and it is never covered by
a horizon as it will be shown below.

%
\subsection{``Astrophysical'' Conditions}
%
Besides the energy conditions, we consider three additional physical
plausible conditions \cite{Kolassis88} which allow us to associate the
source studied here with a stellar object. These conditions are
\bq
\lb{3.21}
\frac{\partial \rho}{\partial \R} < 0,
\; \; \; \; \; \; \; \; \; \; \; \;
p > 0,
\; \; \; \; \; \; \; \; \; \; \; \;
\frac{\partial p}{\partial \R} < 0,
\; \; \; \;  \; \; \; \; \; \; \;
\frac{\partial \left( q_\alpha q^\alpha\right)}{\partial \R} > 0.
\; \; \; \;  \; \; \; \; \; \; \;
\eq

The graphics in the figure (\ref{rho_pr_q}) show clearly that
these conditions are fulfilled.

\begin{figure}[]
\begin{center}
\leavevmode
\psfig{file=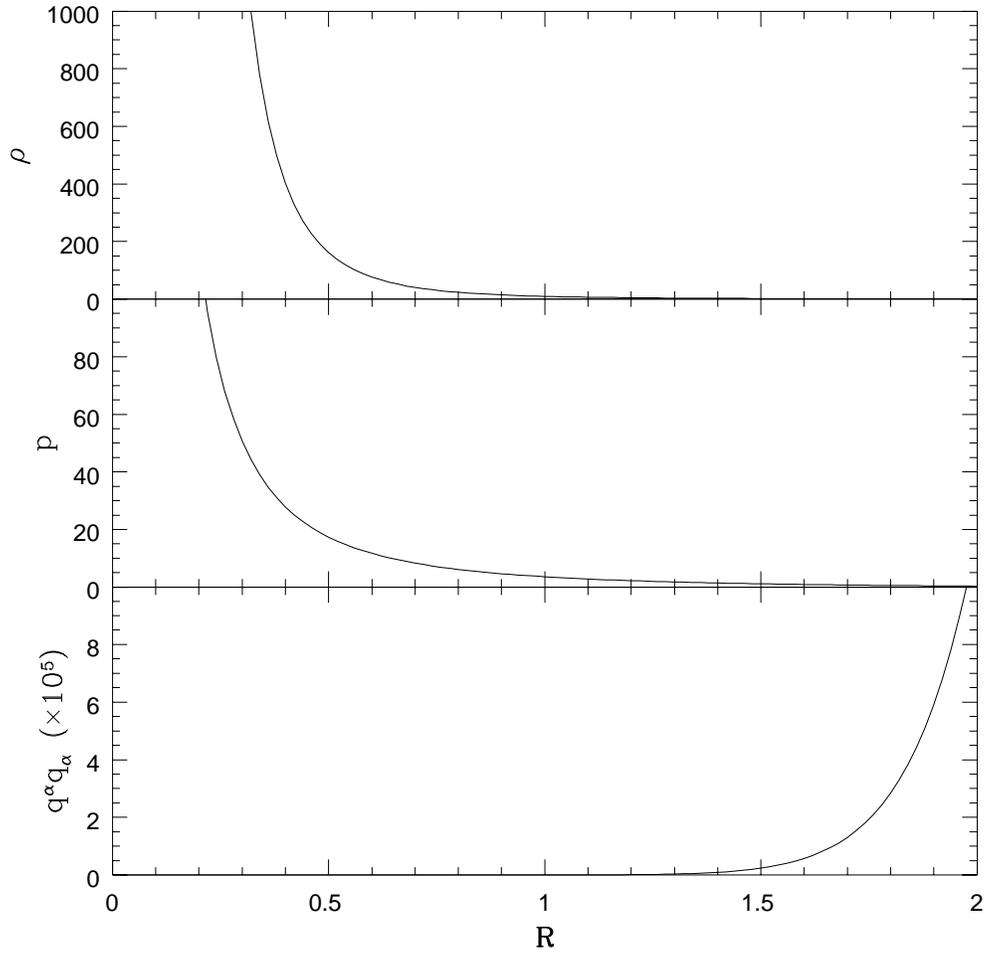,width=1.0\textwidth,angle=0}
\caption{The ``astrophysical'' conditions ($c=1/15$, $t=-1$).
The geometrical radius $\R$ is in units of second, $\rho$
and $p$ are in units of sec$^{-2}$ and $q^\alpha
q_\alpha$ is in units of sec$^{-4}$
}
\label{rho_pr_q}
\end{center}
\end{figure}

%

\begin{figure}[]
\begin{center}
\leavevmode
\psfig{file=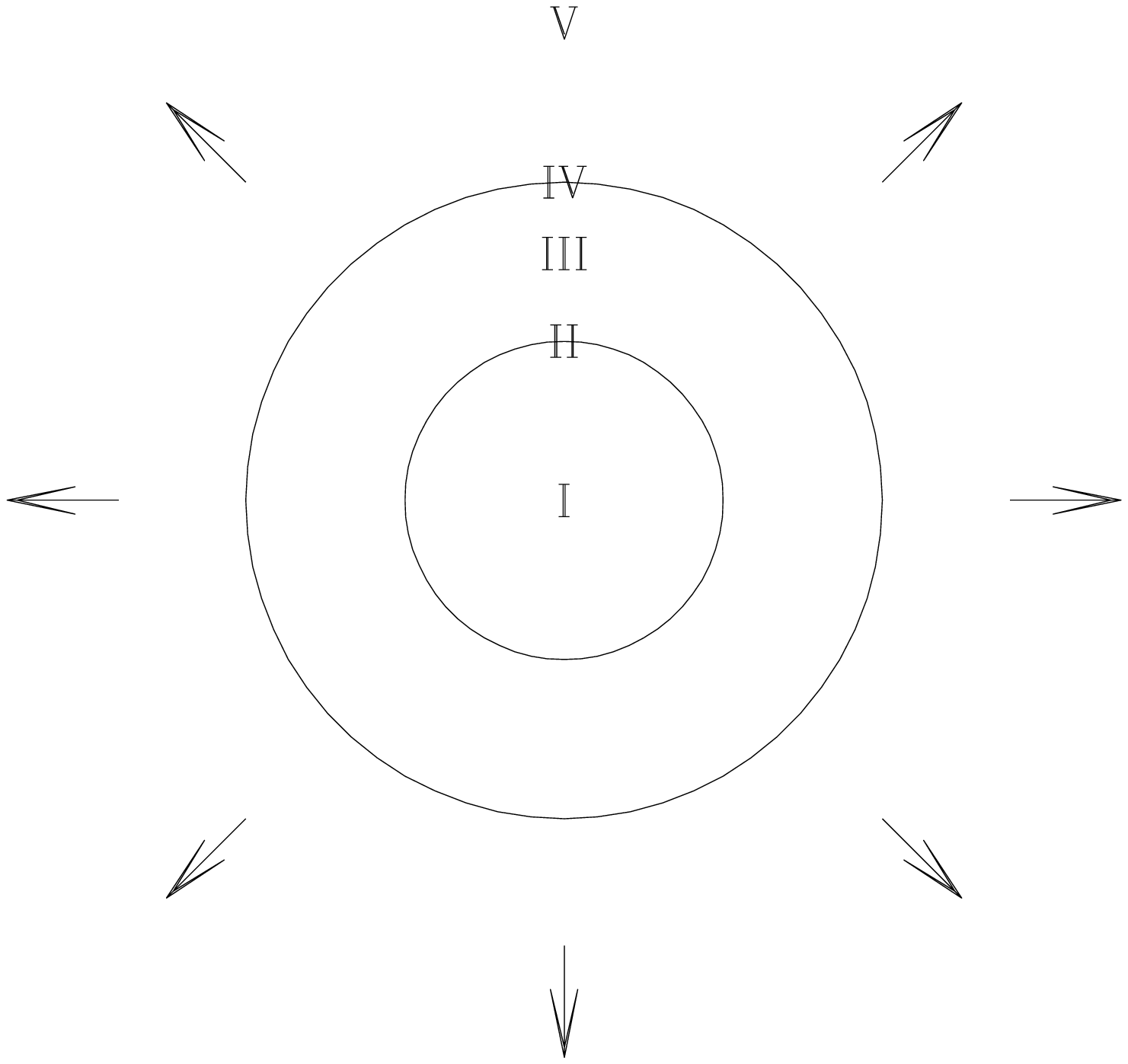,width=1.0\textwidth,angle=0}
\caption{A scheme for the thick shell.
The fluid with a heat flow fulfills the region III. The region I
is the vacuum Minkowski spacetime inside the shell. Region
V is the null radiation fluid (Vaidya's spacetime) outside the shell.
The regions II and IV represent the $\Sigma_1$ and $\Sigma_2$
hypersurfaces, respectively.}
\label{junction}
\end{center}
\end{figure}

\section{Matching Regions I and III}

In this section we will present the junction conditions for the regions I
(Minkowski spacetime) and III (fluid with heat flow) and for the regions III
(fluid with heat flow) and V (Vaidya's spacetime), as shown in the
figure \ref{junction}.

\subsection{Junction Conditions for the region II }

The Minkowski metric can be written as

\begin{equation}
ds^2_{I}=dV^2-dR^2-R^2d\theta^2-R^2\sin^2 \theta d\phi^2,
\label{eq:ds2I}
\end{equation}
and the shell fluid can be written as
\begin{equation}
ds^2_{III}=A^2(r,t) \left[ dt^2-dr^2-r^2d\theta^2-r^2\sin^2
\theta d\phi^2 \right].
\label{eq:ds2III}
\end{equation}

The metric of the junction hypersurface $\Sigma_1$ is given by
\begin{equation}
ds^2_{II}=d\tau^2-\beta^2(\tau)(d\theta^2+\sin^2 \theta d\phi^2),
\label{eq:ds2II}
\end{equation}
where the time coordinate $\tau$ is defined only on $\Sigma_1$.

Thus, using the first fundamental forms on $\Sigma_1$ we have
\begin{equation}
(ds^2_{I})_{\Sigma_1}=(ds^2_{II})_{\Sigma_1}=(ds^2_{III})_{\Sigma_1},
\label{eq:junction1}
\end{equation}
resulting
\begin{equation}
\left[1-\left({{dR} \over {dV}}\right)^2_{\Sigma_1}\right]dV^2 = d\tau^2 =
A^2(r_{\Sigma_1},t)dt^2,
\label{eq:junction2}
\end{equation}
and
\begin{equation}
R_{\Sigma_1}(V) = \beta(\tau) = A(r_{\Sigma1},t)r_{\Sigma_1}.
\label{eq:junction3}
\end{equation}

The extrinsic curvature is given by
\begin{equation}
K_{ab}=-n_{\alpha}
{{\partial^2x^{\alpha}} \over {\partial \xi^a \partial \xi^b}}
-n_{\alpha}\Gamma^{\alpha}_{\beta \gamma}
{{\partial x^{\beta}}  \over {\partial \xi^a}}
{{\partial x^{\gamma}} \over {\partial \xi^b}},
\label{eq:kab}
\end{equation}
where $n_\alpha$ is the unit normal vector, $x^\alpha$ refers to the
equation of $\Sigma_1$ and $\xi^a$ take the values $\tau$, $\theta$ and
$\phi$.

We can easily show that, using the Lichnerowicz \cite{Lichnerowicz} and
O'Brien and
Synge \cite{OBrien} junction conditions, it is impossible to match a
spacetime with
heat flow and a Minkowski spacetime.
This can be due to the continuity of the radial flux of momentum across
the hypersurface $\Sigma_1$ (see \cite{Bonnor}, page 283).
Since we cannot make a smooth matching without a physical surface layer
between
them, we must introduce a thin
shell with a energy-momentum tensor $T_{ab}$.  Following Israel
\cite{Israel66},
the relation between
the energy-momentum tensor of the layer and the extrinsic curvature
(equation \ref{eq:kab}) of the spacetime is
\begin{equation}
[K^I_{ab}-K^{III}_{ab}]_{\Sigma_1}=-\kappa(T_{ab} - g_{ab}T/2),
\label{eq:secform1}
\end{equation}
where $K^I_{ab}$ is the extrinsic curvature of the spacetime I and
$K^{III}_{ab}$
is the extrinsic curvature of the spacetime III.  The indices take the
values
$\tau$, $\theta$ and $\phi$, and
\begin{equation}
T =g^{ab}T_{ab}.
\label{eq:secform2}
\end{equation}

The unit normal vectors of the spacetimes I and III can be written as
\begin{equation}
n^{I}_{\alpha}=(-R^*_{\Sigma_1}, V^*, 0, 0),
\label{eq:norm1}
\end{equation}
and
\begin{equation}
n^{III}_{\alpha}=(0, A(r_{\Sigma_1},t), 0, 0),
\label{eq:norm2}
\end{equation}
where the symbol $*$ denotes differentiation with respect to the coordinate
$\tau$.

Using the extrinsic curvature (equation \ref{eq:kab}) and the metrics
(equations \ref{eq:ds2I} and \ref{eq:ds2III}) and the equation
$A^2(r,t)=t^2/(ct^2+r^2)$ we have
\begin{equation}
K^I_{\tau \tau}=[R^*V^{**}-V^*R^{**}]_{\Sigma_1},
\label{eq:ktautauI}
\end{equation}
\begin{equation}
K^I_{\theta \theta}=[V^{*}R]_{\Sigma_1},
\label{eq:kthetathetaI}
\end{equation}
\begin{equation}
K^I_{\phi \phi}=[V^{*}R\sin^2 \theta]_{\Sigma_1},
\label{eq:kphiphiI}
\end{equation}
\begin{equation}
K^{III}_{\tau \tau}=\left[{{2r(t^*)^2t^2} \over
{(r^2+ct^2)^2}}\right]_{\Sigma_1},
\label{eq:ktautauIII}
\end{equation}
\begin{equation}
K^{III}_{\theta \theta}=\left[{{rt^2(ct^2-r^2)} \over
{(r^2+ct^2)^2}}\right]_{\Sigma_1},
\label{eq:kthetathetaIII}
\end{equation}
and
\begin{equation}
K^{III}_{\phi \phi}=\left[{{rt^2(ct^2-r^2)} \over {(r^2+ct^2)^2}}\sin^2
\theta\right]_{\Sigma_1}.
\label{eq:kphiphiIII}
\end{equation}

Thus, substituting equations (\ref{eq:ktautauI})-(\ref{eq:kphiphiIII}) into
equation (\ref{eq:secform1}) we get
\begin{equation}
T_{\tau \tau} = {1 \over \kappa} \left[ {2V^* \over R} - {{2(ct^2-r^2)}
\over {rt^2}} \right]_{\Sigma_1},
\label{eq:ttautau}
\end{equation}
\begin{equation}
T_{\theta \theta} = {1 \over \kappa} \left[
R^2(R^*V^{**}-V^*R^{**})-RV^*+{{rt^2(ct^2-3r^2)} \over {(ct^2+r^2)}}
\right]_{\Sigma_1},
\label{eq:tthetatheta}
\end{equation}
and
\begin{equation}
T_{\phi \phi} = T_{\theta \theta} \sin^2 \theta.
\label{eq:tphiphi}
\end{equation}

The energy-momentum tensor of a perfect fluid can be constructed as
\begin{equation}
T_{ab} = \sigma u_a u_b + \eta (\Theta_a\Theta_b + \Phi_a \Phi_b),
\label{eq:tab}
\end{equation}
where
\begin{equation}
u_a =\delta^\tau_a,
\label{eq:ua}
\end{equation}
\begin{equation}
\Theta_a=\beta\delta^\theta_a,
\label{eq:thetaa}
\end{equation}
and
\begin{equation}
\Phi_a=\beta\sin \theta \delta^\phi_a.
\label{eq:phia}
\end{equation}

Comparing equations (\ref{eq:ttautau})-(\ref{eq:tphiphi}) and
(\ref{eq:tab}),
we can write that

\begin{equation}
\kappa \sigma
 = \left[{2V^* \over R} - {{2(ct^2-r^2)} \over {rt^2}}\right]_{\Sigma_1},
\label{eq:rho}
\end{equation}
and
\begin{equation}
\kappa \eta
 = \left[(R^*V^{**}-V^*R^{**})-{V^* \over R}+{{ct^2-3r^2} \over
{rt^2}}\right]_{\Sigma_1}.
\label{eq:p}
\end{equation}

The equations (\ref{eq:rho})-(\ref{eq:p}) can be rewritten as
\begin{equation}
\kappa \sigma
 = \left[{{2(1+{R^*}^2)^{1 \over 2}} \over R} - {{2(ct^2-r^2)} \over
{rt^2}}\right]_{\Sigma_1},
\label{eq:rho1}
\end{equation}
and
\begin{equation}
\kappa \eta
 = \left[-{(1+{R^*}^2)^{1 \over 2} \over R}-{{R^{**}} \over {(1+{R^*}^2)^{1
\over 2}}}+{{ct^2-3r^2} \over {rt^2}}\right]_{\Sigma_1}.
\label{eq:p1}
\end{equation}

\subsection{Energy Conditions for the Thin Shell}

Differentiating the equation (\ref{eq:junction3}) with respect to $\tau$ we
get
\begin{equation}
R^*_{\Sigma_1}=\left[{{2r^3} \over {t(ct^2+r^2)}}\right]_{\Sigma_1}.
\label{eq:rstau}
\end{equation}
Thus, we can see that the hypersurface $\Sigma_1$ collapses since $t<0$ thus
$R^*<0$.

The weak energy conditions are given by $\sigma
 \ge 0$ and $\sigma+\eta \ge 0$, thus from
the first condition we get
\begin{equation}
\left\{ ct^2+r^2\left[{1 + {{4r^6} \over {t^2(ct^2+r^2)^2}}}\right]^{1 \over
2}\right\}_{\Sigma_1} \ge (ct^2-r^2)_{\Sigma_1},
\label{eq:wc1}
\end{equation}
which is always satisfied.

>From the second condition we have

\begin{equation}
\left[{R^{**} \over {(1+{R^*}^2)^{1 \over 2}}}\right]_{\Sigma_1} \le
\left[{{1+{R^*}^2} \over {R(1+{R^*}^2)^{1 \over 2}}} - {1 \over
R}\right]_{\Sigma_1} = F_1
\label{eq:wc2}
\end{equation}

In order to hold the dominant energy conditions we must have the conditions
$\sigma \ge 0$ and $\sigma+ \eta \ge 0$ and $\sigma-\eta \ge 0$.  From this
last condition
we have

\begin{equation}
\left[{R^{**} \over {(1+{R^*}^2)^{1 \over 2}}}\right]_{\Sigma_1} \ge
\left[-{3{(1+{R^*}^2)^{1 \over 2}} \over R} + {{3ct^2-5r^2} \over
{rt^2}}\right]_{\Sigma_1} = F_2
\label{eq:dc1}
\end{equation}

In order to hold the strong energy conditions we must have the conditions
$\sigma+\eta \ge 0$ and $\sigma+2\eta \ge 0$.  From this last condition we
have

\begin{equation}
\left[{R^{**} \over {(1+{R^*}^2)^{1 \over 2}}}\right]_{\Sigma_1} \le \left[-
{{2r} \over {t^2}}\right]_{\Sigma_1} = F_3.
\label{eq:sc1}
\end{equation}
Thus, we can conclude that $R^{**} \le 0$.  It is easy to show that $F_1 \ge
0$ and that
equation (\ref{eq:wc2}) is always satisfied.

Finally, we can show that $F_2 < F_3$.  Thus, in order to hold all the
energy conditions
we must have

\begin{equation}
F_2 \le \left[{R^{**} \over {(1+{R^*}^2)^{1 \over 2}}}\right]_{\Sigma_1} \le
F_3.
\label{eq:aec}
\end{equation}

The above equation implies in two conditions, that can be written in terms
of the self-similar variable $z$, given by
\bqn
\lb{eq:aeca}
\left[ z^8+(6c-4)z^6+\left(9c^2-1\right)z^4-2cz^2-c^2
\right]_{\Sigma_1}&<& 0, \\
\left[ -2z^6+(6c-1)z^4-2z^2-1
+\left(2c-5z^2\right)\sqrt{\left(c+z^2\right)^2
+4z^6} \right]_{\Sigma_1} &<& 0.
\lb{eq:aecb}
\eqn
For the entire
range of $z$ and $c$ imposed by the energy conditions of the thick shell, we
can
see that they are all satisfied. Thus, the Minkowski spacetime can be
identified  as the region above the $x_2(c)$ (the condition energy
frontier), for example, in the figures (\ref{raiz6ahv}) and
(\ref{jvaifinal}).

\section{Matching Regions III and V}

In this section
we present and analyze the junction conditions for regions
III and V (Vaidya's spacetime \cite{Vai}), as shown in the
figure \ref{junction}.

\subsection{Junction Conditions for the region IV}

The Vaidya's metric is given by
\begin{equation}
ds^2_{V}=[1-2m({\bf v})/{\bf r}]d{\bf v}^2-2d{\bf r}d{\bf v}-{\bf
r}^2d\theta^2-{\bf r}^2\sin^2 \theta d\phi^2,
\label{eq:ds2V}
\end{equation}
where $m({\bf v})$ is an arbitrary function the time ${\bf v}$.

The metric of the hypersurface $\Sigma_2$ is given by
\begin{equation}
ds^2_{IV}=dv^2-\gamma^2(v)(d\theta^2+\sin^2 \theta d\phi^2).
\label{eq:ds2III-V}
\end{equation}

>From the junction condition

\begin{equation}
(ds^2_{III})_{\Sigma_2}=(ds^2_{IV})_{\Sigma_2}=(ds^2_{V})_{\Sigma_2},
\label{eq:junction4}
\end{equation}
we obtain
\begin{equation}
{dt \over {d v}}=A(t,r_{\Sigma_2})^{-1},
\label{eq:ts}
\end{equation}
\begin{equation}
A(t, r_{\Sigma_2})r_{\Sigma_2}=\gamma(v)={{\bf r}_{\Sigma_2}({\bf v})},
\label{eq:cs}
\end{equation}
and
\begin{equation}
\left( d{\bf v} \over {d v} \right)^{-2}_{\Sigma_2}=\left( 1 - {2m \over
{\bf r}} +
2 {d{\bf r} \over d{\bf v}} \right)_{\Sigma_2},
\label{eq:dvdv}
\end{equation}
where $v$ is a time coordinate defined only on $\Sigma_2$.

The unit normal vectors to $\Sigma_2$
(for details see \cite{Santos85}) are given by
\begin{equation}
n^{III}_{\alpha}=A(t, r_{\Sigma_2})\delta^1_{\alpha},
\label{eq:ni}
\end{equation}

\begin{equation}
n^{V}_{\alpha}=\left( 1 - {2m \over {\bf r}} +
2 {d{\bf r} \over d{\bf v}} \right)^{-1/2}_{\Sigma_2}
\left( -{{d{\bf r}} \over d{\bf v}}\delta^0_{\alpha} +
\delta^1_{\alpha} \right)_{\Sigma_2}.
\label{eq:no}
\end{equation}

The non-vanishing extrinsic curvature components are given by

\begin{equation}
K^{III}_{v v}=-\left[ {\left( {dt \over {d {v}}} \right)}^2 {A'}
\right]_{\Sigma_2},
\label{eq:k00i}
\end{equation}
\begin{equation}
K^{III}_{\theta \theta}= \left[ {r(Ar)'} \right]_{\Sigma_2},
\label{eq:k22i}
\end{equation}
\begin{equation}
K^{III}_{\phi \phi}=K^{III}_{\theta \theta} \sin^2 \theta,
\label{eq:k33i}
\end{equation}
\begin{equation}
K^{V}_{v v}=\left[ {d^2{\bf v} \over {d {v}^2}}
{\left( d{\bf v} \over d {v} \right)}^{-1}
-{\left( d{\bf v} \over d {v} \right)} {m \over {{\bf r}^2}}
\right]_{\Sigma_2},
\label{eq:k00o}
\end{equation}
\begin{equation}
K^{V}_{\theta \theta}= \left[ {\left( d{\bf v} \over d {v} \right)}
\left( 1 - {2m \over {\bf r}} \right){\bf r}+{d{\bf r} \over d {v} }{\bf r}
\right]_{\Sigma_2},
\label{eq:k22o}
\end{equation}
\begin{equation}
K^{V}_{\phi \phi}=K^{V}_{\theta \theta} \sin^2 \theta,
\label{eq:k33o}
\end{equation}
where the prime denotes differentiation with respect to the coordinate $r$.

>From the equations (\ref{eq:k22i}) and (\ref{eq:k22o}) we have

\begin{equation}
\left[ {\left( d{\bf v} \over d {v} \right)}
\left( 1 - {2m \over {\bf r}} \right){\bf r}+{d{\bf r} \over d {v} }{\bf r}
\right]_{\Sigma_2}= \left[ {r(Ar)'} \right]_{\Sigma_2}.
\label{eq:k22ik22o}
\end{equation}

With the help of equations (\ref{eq:ts}), (\ref{eq:cs}), (\ref{eq:dvdv}),
we can write (\ref{eq:k22ik22o}) as

\begin{equation}
m=\left\{ {{rA} \over 2}\left[ 1 + {\left( {r\dot A} \over A \right)}^2 -
{\left( 1+{rA' \over A} \right)}^2 \right] \right\}_{\Sigma_2},
\label{eq:ms}
\end{equation}
which is the total energy entrapped inside the hypersurface $\Sigma_2$
\cite{CahMac70} and
where the dot denotes differentiation with respect to the coordinate $t$.

Using equations (\ref{eq:k00i}) and (\ref{eq:k00o}) and following the same
procedure
described in \cite{Bonnor}, we can finally write

\begin{equation}
\label{eq:pqbs}
(p)_{\Sigma_2}=(qA)_{\Sigma_2}.
\end{equation}

\begin{figure}[]
\begin{center}
\leavevmode
\psfig{file=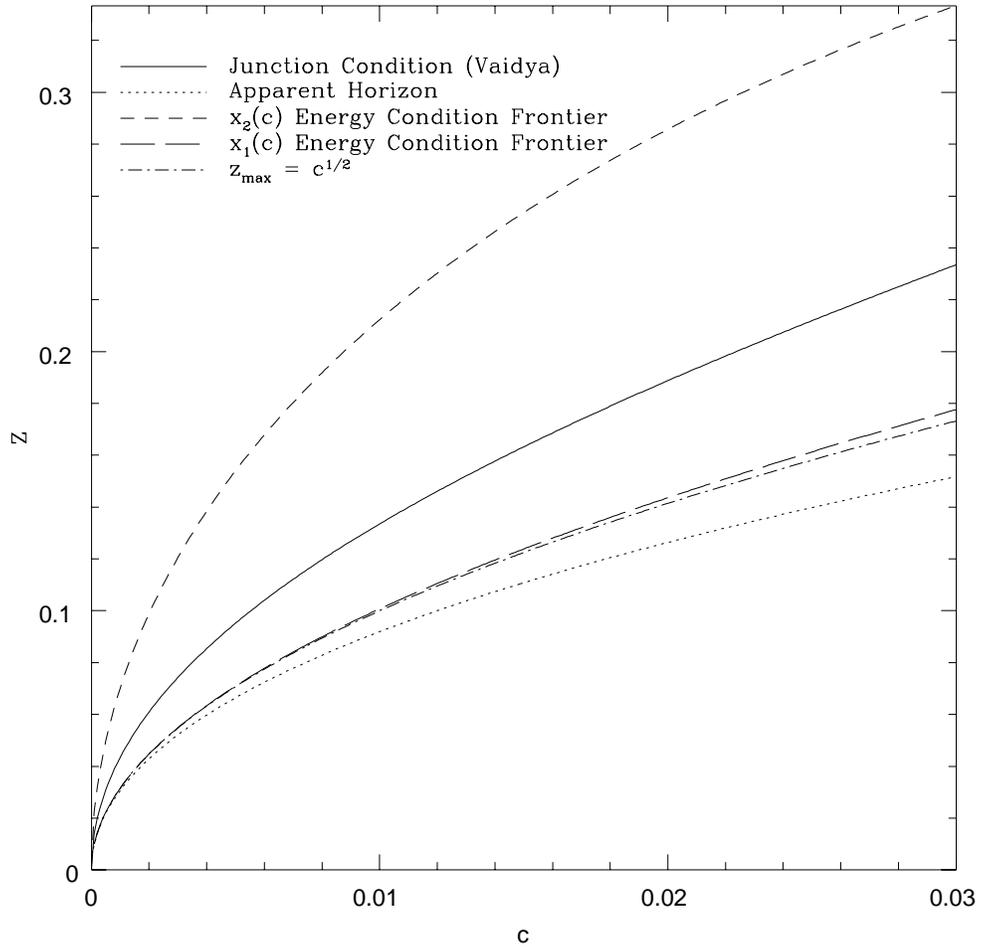,width=1.0\textwidth,angle=0}
\caption{The energy condition frontiers, apparent horizon and junction
condition, in a $c$-$z$ diagram.}
\label{raiz6ahv}
\end{center}
\end{figure}

\begin{figure}[]
\begin{center}
\leavevmode
\psfig{file=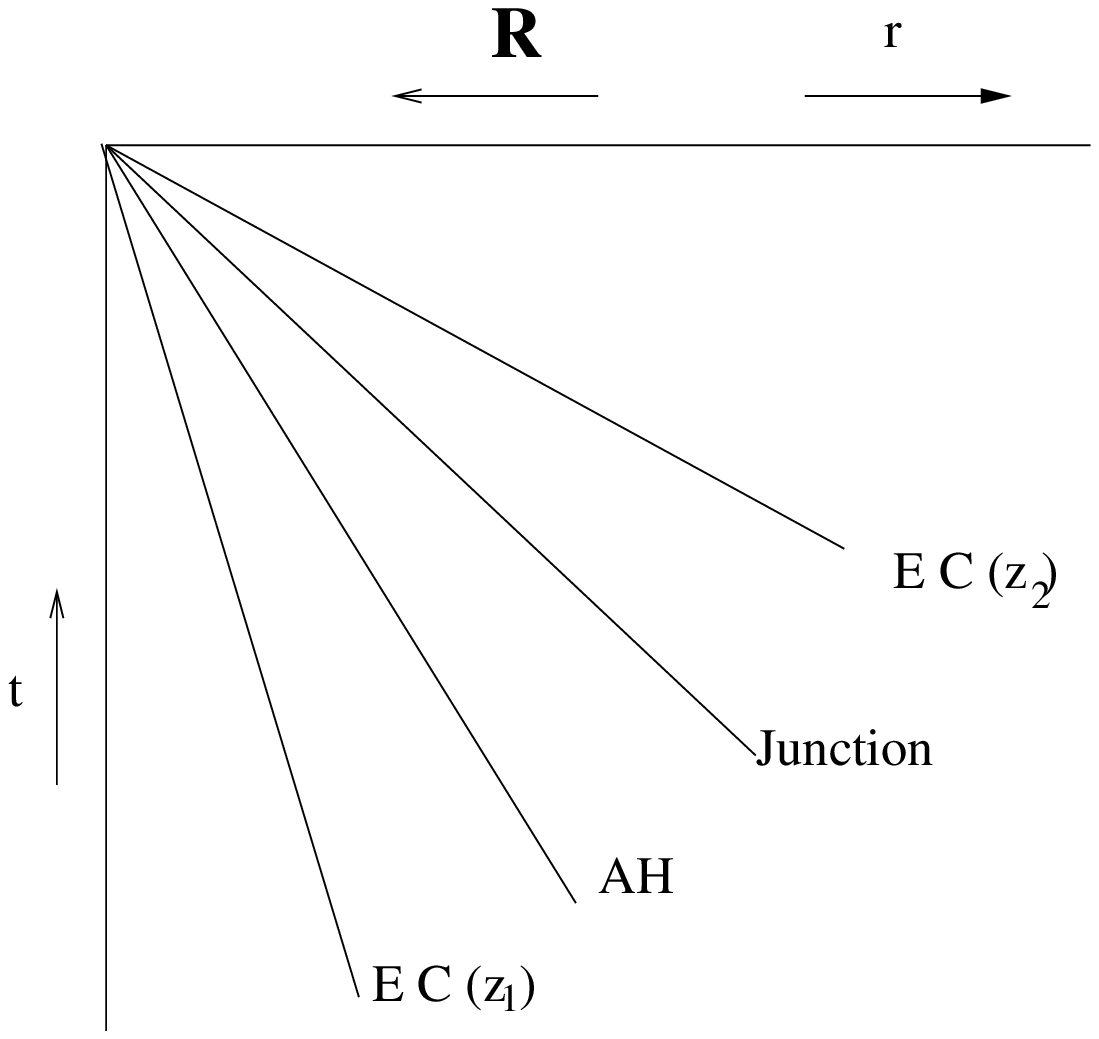,width=1.0\textwidth,angle=0}
\caption{The energy condition frontiers, apparent horizon and junction
condition, in the t-$\R$ diagram. The energy frontiers
$z_1$ and $z_2$ correspond to $x_1$ and $x_2$, respectively. EC denotes
energy conditions and AH denotes apparent horizon. } \label{jvaifinal}
\end{center}
\end{figure}

Substituting the equations for $p$ and $q$ into (\ref{eq:pqbs}) we get
(see figures (\ref{raiz6ahv}) and (\ref{jvaifinal}))

\begin{equation}
2z^3_{\Sigma_2}+(1+3c)z^2_{\Sigma_2}+2cz_{\Sigma_2}-2c=0.
\label{eq:pqbs1}
\end{equation}

The hypersurface of the matching is always inside the
interval allowed by the energy conditions of the thick shell.

\subsection{Inexistence of an Apparent Horizon}

Below we prove that the singularity at $t$ = 0 and \R = 0 is naked
for the thick shell.

Its is known that on the apparent horizon $g_{tt}$ changes its sign.
We say that a black hole is formed during a collapse process when its
boundary hypersurface cross inwards an apparent horizon.  Since at the
initial stage $g_{tt}$ is positive, the condition for the inexistence of
an apparent horizon is
\bq
\lb{eq:ahvay}
\left(1 - \frac{2 m}{\R} \right)_{\Sigma_2} > 0.
\eq

Considering the matching with Vaidya's solution we can rewrite equation
(\ref{eq:ahvay}) as
\bq
\lb{eq:ahvay1}
- \left[ \frac{(1-z)(z^2+c)z}{(1-z^2)} \right]_{\Sigma_2} < \frac{1}{2}.
\eq
\noindent

Since the ranges of $c$ and $z_{\Sigma}$ are $0\leq c \leq 1/15$ and
$0 \leq z_{\Sigma} < 1/3$, respectively, it is easy to verify that this
inequality is always satisfied.

\section{Conclusions}

In this paper, we have studied a complete class of solutions,
(shear-free, conformally flat, self-similar) which represents
gravitational collapse of a fluid with heat flow. The fluid
satisfies all the energy conditions only in the region $ z_{1} < z
< z_{2}$, where $z$ denotes the self-similar variable. Thus, to
have a physically realistic model, one needs to do some kind of
``surgery", that is, cutting the spacetime along the two
hypersurfaces $z = z_{1,2}$ and then matching the part $ z_{1} < z
< z_{2}$ to some other regions of spacetime. Since these
hypersurfaces are time-like, one can always be able to do so.
Then, the resultant spacetime will represent gravitational
collapse of a thick spherical shell. However, if we do not
consider the strong energy condition, we can have a collapse of a
full spherical body, instead of a thick shell. It has been shown
that, at least for the self-similarity of first kind,  the thick
shell collapses always forming naked singularities.

This, together with the results obtained in \cite{deutsche,ind02}
show that the results obtained from the gravitational collapse of
a perfect fluid \cite{JDM01} cannot be generalized to anisotropic
fluid, and the formation of naked singularities depends not only
on the shear of the fluid, but also on pressures and heat flow.

\section*{Acknowledgments}

We would like very much to thank A. Z. Wang for valuable discussions and
suggestions. The financial assistance from UERJ/FAPERJ (MFAdaS) and CNPq
(JFVR) is gratefully acknowledged.

\end{document}